\documentclass{webofc}
\usepackage[varg]{txfonts}   
\begin{document}
\title{Revealing the Origin of Mass through Studies of Hadron Spectra and Structure}
%
%

\author{\firstname{} \lastname{\emph{Craig} \emph{D}. Roberts}\inst{1,2}\fnsep\thanks{\email{cdroberts@nju.edu.cn}}}


\institute{School of Physics, Nanjing University, Nanjing, Jiangsu 210093, China
\and
           Institute for Nonperturbative Physics, Nanjing University, Nanjing, Jiangsu 210093, China
          }

\abstract{%
The Higgs boson is responsible for roughly 1\% of the visible mass in the Universe.  Obviously, therefore, Nature has another, very effective way of generating mass.  In working toward identifying the mechanism, contemporary strong interaction theory has arrived at a body of basic predictions, \emph{viz}.\ the emergence of a nonzero gluon mass-scale, a process-independent effective charge, and dressed-quarks with constituent-like masses.  These three phenomena -- the pillars of emergent hadron mass (EHM) -- explain the origin of the vast bulk of visible mass in the Universe.  Their expressions in hadron observables are manifold.  This contribution highlights a few; namely, some of the roles of EHM in building the meson spectrum, producing the leading-twist pion distribution amplitude, and moulding hadron charge and mass distributions.
}
\maketitle
\section{Introduction}
\label{intro}

The existence and character of our Universe depends critically on, \emph{inter alia}, the following empirical facts.
(\emph{A}) The proton is massive, \emph{i.e}., the $m_p\approx 1\,GeV$ proton mass scale associated with strong interactions is vastly different from that which characterises electromagnetism, the electron mass $m_e = m_p/1836$.
(\emph{B}) The proton is stable; at least, the lower bound on its lifetime is $10^{25}$-times greater than the age of our Universe.
(\emph{C}) The pion is unnaturally light.  It is not massless; yet, it has a lepton-like mass, despite being a strongly interacting composite object built from the same sort of valence degrees-of-freedom as the massive, stable proton.
These features of our Universe are examples of what may be called emergent phenomena within the Standard Model (SM) of particle physics: low-level rules producing high-level phenomena with apparently remarkable complexity.

The SM has one phenomenologically understood mass-generating mechanism; namely, that tied to the Higgs boson (HB).  Its impacts are crucial to the evolution of our Universe,  However, alone, HB couplings into quantum chromodynamics (QCD) are responsible for only $\sim 1$\% of $m_p$.  Evidently, Nature has another, very effective mechanism for producing mass, which is now identified as emergent hadron mass (EHM).  A detailed picture of the proton mass budget is drawn, \emph{e.g}., in Ref.\,\cite[Fig.\,1A]{Ding:2022ows}: EHM is the source of 94\% of $m_p$.  The remaining 5\% is generated by constructive EHM+HB interference.  The $\rho$-meson mass budget looks much like that of the proton -- see Ref.\,\cite[Fig.\,1B]{Ding:2022ows}.  However those of the pion and kaon are very different -- see Ref.\,\cite[Figs.\,1C, 1D]{Ding:2022ows}.  Indeed, without HB effects, the $\pi$ and $K$ would be indistinguishable, massless Nambu-Goldstone bosons \cite{Horn:2016rip}.

These observations raise many questions.  For instance: what is EHM and can it be explained by QCD; (how) is EHM connected with gluon and quark confinement within hadrons and dynamical chiral symmetry breaking (DCSB); and what is the role of the Higgs in modulating the observable properties of hadrons?  Overall, what is and wherefrom mass?

A modern understanding began roughly forty years ago, when it was realised that, owing to their self interactions, QCD gluons dynamically generate a mass for themselves \cite{Cornwall:1981zr}.  Since then, this prediction from continuum analyses of QCD has been refined, as summarised elsewhere \cite{Binosi:2022djx, Ferreira:2023fva},  and confirmed in numerical simulations of the lattice regularised theory \cite{Ayala:2012pb}.  Today, therefore, it is a theoretical fact that strong gluon self-interactions transform massless gluon partons into gluon quasiparticles, each of which is characterised by a momentum-dependent mass function that is large at infrared momenta -- see Ref.\,\cite[Fig.\,2]{Ding:2022ows}.  This is truly \emph{mass from nothing}: an interacting theory, written in terms of massless gluon partons, produces massive dressed gluon fields.

Exploiting the emergence of a gluon mass, it becomes possible to rigorously define and calculate a unique, process-independent QCD analogue of the Gell-Mann--Low effective charge in QED \cite{GellMann:1954fq}.  As in QED, the running of this charge is entirely determined by the momentum dependence of the gauge boson (gluon, in QCD) vacuum polarisation \cite{Cui:2019dwv}.  Featured, \emph{e.g}., in Ref.\,\cite[Fig.\,4.1]{Deur:2023dzcX}, the momentum dependence of this running charge matches that of perturbative QCD (pQCD) as the probe momentum decreases from the ultraviolet toward $m_p$.  However, unlike the pQCD charge, it does not diverge at $\Lambda_{\rm QCD}$: there is no Landau pole \cite[Sec.\,2]{Deur:2023dzcX}.  Instead, a qualitative change occurs in the neighbourhood of the gluon mass ($\approx m_p/2$).  That mass eliminates antiscreening, so the charge stops running and QCD becomes, once again, practically conformal, \emph{viz}.\ interactions become scale independent, just as they were in the Lagrangian.  With such a running coupling, the ``infrared slavery'' picture of confinement is seen to be false: there is no linear potential between dynamical colour sources.  Confinement is otherwise realised in QCD \cite[Sec.\,5]{Ding:2022ows}.

Introducing these features of QCD's gauge sector into the quark gap equation, it is found that massless quark partons, too, are transmogrified into quasiparticles with a momentum dependent mass that is large at infrared momenta -- see Refs.\,\cite[Fig.\,2]{Ding:2022ows}, \cite[Fig.\,2.5]{Roberts:2021nhw}.

This body of remarks introduces the three pillars of EHM:
(\emph{i}) emergence of a nonzero gluon mass-scale;
(\emph{ii}) a process-independent effective charge, which saturates at infrared momenta;
and (\emph{iii}) dressed-quarks with constituent-like masses in the infrared but current masses in the ultraviolet.
EHM is expressed in every strong interaction observable.  Theory is now challenged to identify its measurable consequences and experiment with testing the predictions so that EHM might come to be understood and the boundaries of the SM may finally be drawn.  On these scores, there is room for optimism, with new facilities in operation, under construction, and in planning \cite{Brodsky:2020vco, Chen:2020ijn, Anderle:2021wcy, Arrington:2021biu, Wang:2022xad, Accardi:2023chbP, Carman:2023zke, Quintans:2022utc}.

\section{Spectrum of Mesons with Strangeness}
\label{SecSpectrum}
Regarding the spectrum of hadrons, results from quark models are still often cited as benchmarks, \emph{e.g}., Ref.\,\cite[PDG]{Workman:2022ynf} states: ``The spectrum of baryons and mesons exhibits a high degree of regularity.  The organizational principle which best categorizes this regularity is encoded in the quark model.  All descriptions of strongly interacting states use the language of the quark model.''  Moreover, it was long ago claimed that \cite{Godfrey:1985xj}: ``\ldots all mesons -- from the $\pi$ to the $\Upsilon$ -- can be described in a unified quark model \ldots''.  These persistent beliefs are challenged by the following facts: neither the so-called quarks nor the potentials in quark models have been shown to possess any mathematical link with QCD -- rigorous or otherwise; the orbital angular momentum and spin used to label quark model states are not Poincar\'e-invariant (not observable) quantum numbers; quark model formulations break many symmetries known to be critical to hadron spectra; and there is no context in which quark models can be systematically improved.  Practically, whilst quark models can be tuned to fit any given spectrum, one should cautious in drawing any insights from such descriptions. 

A systematic approach to continuum bound-state problems in QCD was introduced almost thirty years ago \cite{Munczek:1994zz, Bender:1996bb}.  Amongst other things: the scheme highlighted the importance of preserving continuous and discrete symmetries when formulating bound-state problems; enabled proof of Goldberger-Treiman identities and the Gell-Mann--Oakes--Renner relation \cite{Maris:1997hd}; and opened the door to symmetry-preserving, Poincar\'e-invariant predictions of hadron observables \cite{Maris:2003vk, Eichmann:2016yit, Qin:2020rad}.  The leading-order (rainbow-ladder, RL) term in this truncation scheme works well for ground state hadrons that possess little rest-frame orbital angular momentum between their valence constituents.  However, it is limited by an inability to realistically express impacts of EHM on hadron observables; and this weakness is not overcome at any finite order of elaboration.

Improved schemes, expressing EHM in the kernels of all quantum field equations relevant to hadron bound state problems (\emph{e.g}., Dyson-Schwinger equations - DSEs), have been identified \cite{Chang:2009zb, Chang:2011ei, Binosi:2014aea, Williams:2015cvx, Binosi:2016rxz, Qin:2020jig}.  They work well in applications to ground-state mesons built from $u$, $d$ valence quarks and/or antiquarks, but that is a small subset of the hadron spectrum.  Thus, a recent extension to the spectrum and decay constants of $u$, $d$, $s$ meson ground- and first-excited states is an important step forward \cite{Xu:2022kng}.  EHM is known to generate a significant dressed-quark anomalous chromomagnetic moment (ACM) \cite{Chang:2010hb, Bicudo:1998qb, Bashir:2011dp, Binosi:2016wcx, Kizilersu:2021jen}.  This means that the gluon+quark interaction acquires a new, dynamically generated piece, a phenomenon which can schematically be expressed as follows:
\begin{equation}
\label{EqVertexGap}
\Gamma_\nu^g(q,k)  = \gamma_\nu \to \gamma_\nu + \eta \kappa((k-q)^2)\sigma_{\rho\nu}(k-q)_\rho\,,
\end{equation}
where $\eta \sim 1$ is the strength of the ACM and $\kappa((k-q)^2)$ falls from unity to zero in the same manner as the quark+quark interaction.

\begin{figure}[!t]
\centerline{%
\includegraphics[clip, width=0.85\textwidth]{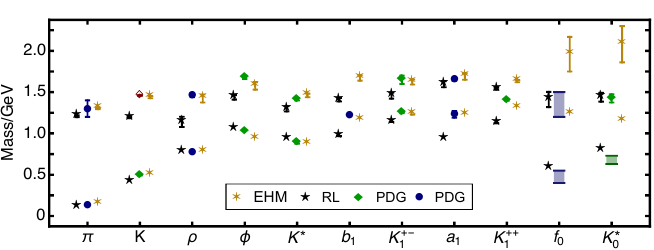}}
\caption{\label{figAllstrange}
Gold six-point stars -- spectrum of low-lying $u$, $d$, $s$ mesons predicted by the bound state kernel developed in Ref.\,\cite{Xu:2022kng};
and black five-point stars -- same spectrum computed using RL truncation.
Comparison spectrum \cite[Summary Tables]{Workman:2022ynf}:
blue circles (bars) -- $u$, $d$ systems; and green diamonds (bar) -- mesons with $s$ and/or $\bar s$ quarks.
Open red diamond -- $K(1460)$, about which little is known.
%
}
\end{figure}

The features and flaws of RL truncation are evident in Fig.\,\ref{figAllstrange}.  Overall, the mean absolute relative difference between RL masses and central experimental values is $13(8)$\%.  This might seem like fair agreement; but there is substantial scatter and there are many qualitative discrepancies, \emph{e.g}., labelling the first excited state with an apostrophe:
$m_{K^\prime} < m_{\pi^\prime}$ in RL truncation, whereas the empirical ordering is opposite, and
the same is true for $(m_{\rho^\prime},m_{\pi^\prime})$,  $(m_{\rho^\prime},m_{K^{\ast \prime}})$;
RL truncation $a_1$-$\rho$ and $b_1$-$\rho$ mass splittings are only one-third of the empirical values because the $b_1$ and $a_1$ mesons are far too light;
$m_{\phi^\prime}-m_\phi$ is half the experimental value;
and the level ordering of the $K_1^{+-}$, $K_1^{++}$ states is incorrect.
In addition, RL truncation binds light quark+antiquark scalar mesons, which are not seen in Nature.

The EHM (ACM) corrected kernel delivers significant improvements.  In fact, the mean absolute relative difference between EHM masses and central experimental values is $2.9(2.7)$\%, which is a 4.6-fold betterment of RL.
Furthermore,
$m_{K^\prime} > m_{\pi^\prime}$,
$m_{\rho^\prime} > m_{\pi^\prime}$,
$m_{\rho^\prime} \approx m_{K^{\ast \prime}}$,
matching empirical results;
the $a_1$-$\rho$ and $b_1$-$\rho$ mass splittings agree well with experiment because including EHM effects in the kernel substantially increased the masses of the $b_1$ and $a_1$ mesons, even while $m_\rho$ was kept fixed;
$m_{\phi^\prime}-m_\phi$ matches experiment to within 2\%;
the level ordering of the $K_1^{+-}$, $K_1^{++}$ states is correct;
and quark+antiquark scalar mesons are heavy, leaving room for necessary final-state interaction contributions \cite{Holl:2005st, Eichmann:2015cra}.

Evidently, both gross features and fine details of the spectrum of $u$, $d$, $s$ mesons are sensitive to expressions of EHM in the gluon+quark interaction.

\section{Wave Functions of Nambu-Goldstone Bosons}
Hadron distribution amplitudes (DAs) may be obtained via a light-front projection of the hadron's Poincar\'e covariant wave function, which itself can be obtained by solving an appropriate set of bound state quantum field equations, such as sketched in Sec.\,\ref{SecSpectrum}.  Possessing a probability density interpretation, these DAs are the nearest thing in quantum field theory to a Schr\"odinger wave function in quantum mechanics.  Consequently, they can be viewed as fundamental to understanding hadron structure, including and especially that of Nature's most fundamental Nambu Goldstone bosons, \emph{viz}.\ $\pi$- and $K$-mesons.

\begin{figure}[!t]
\hspace*{-1ex}\begin{tabular}{ll}
{\sf A} & {\sf B} \\[-2ex]
\includegraphics[clip, width=0.45\textwidth]{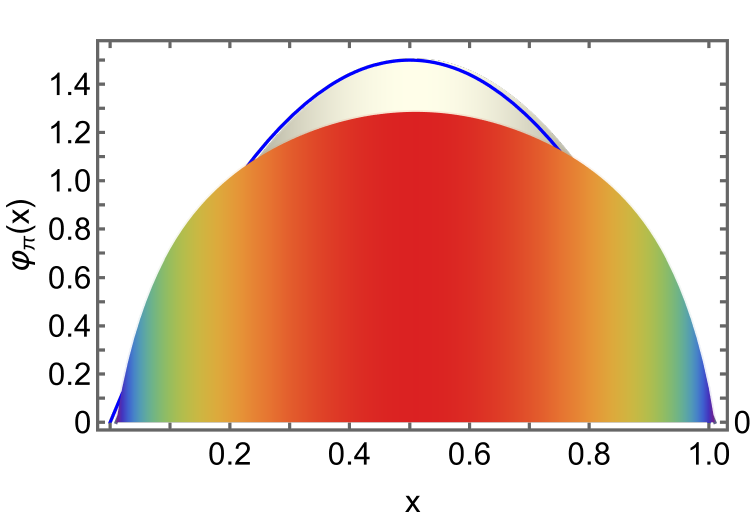} &
\includegraphics[clip, width=0.45\textwidth]{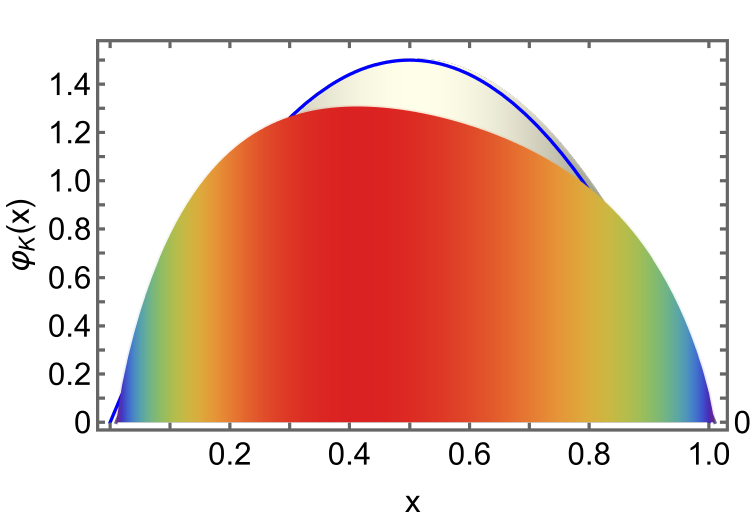}
\end{tabular}
\caption{\label{FigDAs}
\emph{Panel A}.
Pion DA -- dilated colour-shaded function -- compared with $\varphi_{\rm as}(x)$ -- background blue curve.
\emph{Panel B}.
Pion DA -- dilated colour-shaded function -- compared with $\varphi_{\rm as}(x)$.
}
\end{figure}

The form of meson DAs that is valid at asymptotically large energy scales was calculated more than forty years ago \cite{Lepage:1979zb, Efremov:1979qk, Lepage:1980fj}: $\varphi_{\rm as}(x) = 6x(1-x)$.  However, since that time, the $x$ dependence of the pion's leading DA, $\varphi_\pi(x)$, at terrestrially achievable energies has been controversial.  Today, modern theory has converged on a consistent picture \cite{Roberts:2021nhw}.  Namely, relative to $\varphi_{\rm as}(x)$, $\varphi_\pi(x)$ is a dilated convex function -- see Fig.\,\ref{FigDAs} -- whose features are largely determined by the momentum dependence of the dressed-quark mass function, \emph{i.e}., the third pillar of EHM.

In fact, EHM generates dilation in both $\varphi_{\pi,K}$.  The kaon DA is also skewed as a result of the HB generated mass difference between the $\bar s$ and $u$ quark current masses. The ratio of these current quark masses is roughly $25$; so if it were only HB effects driving the distortion, then one might expect to see the $\varphi_{K}$ peak shifted to $x \approx \tfrac{1}{2} \times \tfrac{1}{25} = 0.02$.  However, looking at Fig.\,\ref{FigDAs}B, that expectation is wrong.  Instead, the peak is shifted to $x \approx 0.4 \approx \tfrac{1}{2} \times \tfrac{M_u(0)}{M_{\bar s}(0)}$, where $M_q(0)$ is the dressed-quark constituent-like mass.  Thus, the asymmetry of $\varphi_{K}$ is an expression of interference between EHM and HB mass.  This remains true in heavier systems \cite{Binosi:2018rht}.
These features of meson DAs have widespread impacts in studies of hard processes; \emph{e.g}., the dilation can be verified in measurements of $\pi$ and $K$ elastic form factors at large momentum transfers \cite{Roberts:2021nhw} and in meson+proton Drell-Yan measurements \cite{Brandenburg:1994wf, Xing:2023wuk}.

\section{Empirical Determination of the Pion Mass Distribution}
DAs and parton distribution functions (DFs) provide a one dimensional picture of in-hadron parton properties.  Generalised parton distributions (GPDs), provide an extension of these images to three dimensions.  They add information about the distribution of partons in the plane perpendicular to the bound-state's total momentum, \emph{i.e}., within the light-front itself.  GPDs can be measured in deeply virtual Compton scattering, so long as at least one of the photons possesses large virtuality, and in deeply virtual meson production \cite{Mezrag:2023nkp}.  A key feature of GPDs is that they provide a direct connection between DFs and hadron form factors because any DF may be recovered as the forward limit of the relevant GPD and any elastic form factor can be expressed via a GPD-based sum rule.  Furthermore, and of special importance, a hadron's GPD provides access to its gravitational form factors.  For these reasons, measurements aimed at providing data that will enable GPD extractions are the focus of numerous experimental programmes, either underway or planned \cite{Brodsky:2020vco, Chen:2020ijn, Anderle:2021wcy, Arrington:2021biu, Wang:2022xad, Accardi:2023chbP}.

The expectation value of the energy-momentum tensor in the pion, \emph{viz}.\ the $\pi$ gravitational current, takes the following form:
\begin{equation}
\Lambda^g_{\mu\nu}(P,\Delta) = 2 P_\mu P_\nu \theta_2^\pi(\Delta^2)
 + \tfrac{1}{2}[\Delta^2 \delta_{\mu\nu} - \Delta_\mu \Delta_\nu] \theta_1^\pi(\Delta^2)\,,
\label{Lambdagpi}
\end{equation}
where $P$ is the average momentum of the incoming and outgoing $\pi$, $\Delta$ is the momentum transfer, and $\theta_{2,1}^\pi$ are, respectively, the in-pion mass and pressure distribution form factors.  The pion mass distribution is linked with its GPD as follows:
\begin{equation}
\theta_2^\pi(\Delta^2) = \int_{-1}^1 dx\, 2x\,H_\pi(x,0,-\Delta^2)\,,
\end{equation}
with energy-momentum conservation entailing $\theta_2^\pi(0)=1$.

\begin{figure}[!t]
\leftline{%
\includegraphics[clip, width=0.5\textwidth]{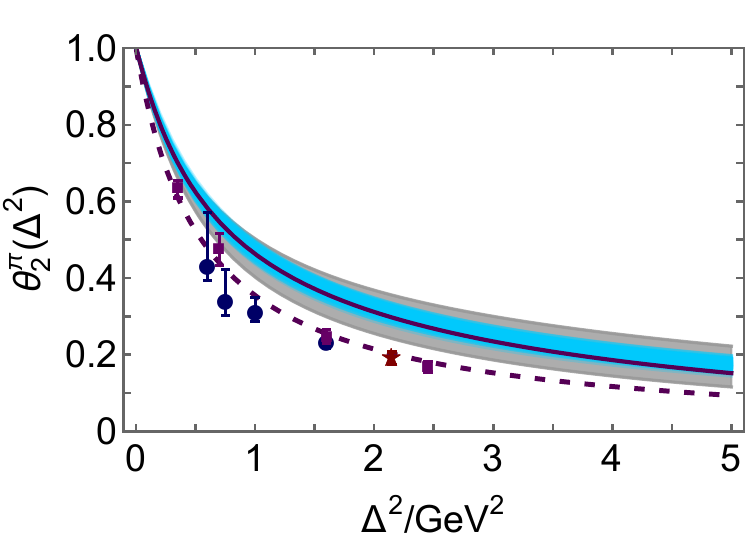}}
\vspace*{-13em}

\rightline{\parbox[c]{0.45\textwidth}{
\caption{\label{Figtheta2}
Pion mass distribution form factor, $\theta_2^\pi(\Delta^2)$ -- Ref.\,\cite{Xu:2023bwv}.
Comparison curves:
CSM prediction for $\theta_2^\pi(\Delta^2)$ in Refs.\,\cite{Zhang:2021mtn, Raya:2021zrz} -- solid purple;
GPD ensemble generated from valence-quark DFs developed in Ref.\,\cite{Cui:2022bxn} using lattice-QCD results \cite{Joo:2019bzr, Sufian:2019bol, Alexandrou:2021mmi} -- grey band.
In addition, each panel displays the CSM prediction for $F_\pi(\Delta^2)$ \cite[Sec.\,4B]{Roberts:2021nhw}, \cite{Chen:2018rwz} -- dashed purple curve.
The data are those for $F_\pi(\Delta^2)$ from Refs.\,\cite{Volmer:2000ek, Horn:2006tm, Tadevosyan:2007yd, Blok:2008jy, Huber:2008id}, included so as to highlight the precision required to distinguish the mass and electromagnetic form factors.
}}}
\end{figure}

Exploiting the connections between GPDs, elastic form factors, and DFs, Ref.\,\cite{Xu:2023bwv} reconstructed the pion GPD from relevant available data  \cite{Conway:1989fs, Amendolia:1984nz, Amendolia:1986wj, Volmer:2000ek, Horn:2006tm, Tadevosyan:2007yd, Blok:2008jy, Huber:2008id}; and, therefrom, the pion mass distribution form factor.  The result is drawn in Fig.\,\ref{Figtheta2}.  Plainly and importantly, $\theta_2(\Delta^2)$ is harder than $F_\pi(\Delta^2)$, \emph{viz}.\ the distribution of mass in the pion is more compact than the distribution of electric charge.  This is an empirical fact.
Indeed, comparing with the pion charge radius \cite{Cui:2022fyr}: $r_\pi = 0.64(2)\,$fm, the data driven prediction is $r_\pi^{\theta_2}/r_\pi = 0.79(3)$.  This translates into a spacetime volume ratio of $0.40(6)$; namely, the pion mass distribution is concentrated within just 40\% of the spacetime volume of the charge distribution.

One may readily understand this empirical fact.  The pion wave function, hence, pion GPD, is independent of the probe.  However, the probe itself focuses on different features of the target constituents.  A target quark carries the same charge, irrespective of its momentum.  So, the pion wave function alone controls the distribution of charge.  On the other hand, the gravitational interaction of a target quark depends on its momentum.  After all, the current is that associated with the energy-momentum tensor.   The pion mass distribution therefore depends on interference between quark momentum growth and wave function momentum suppression as the product $\Delta^2 x^2$ increases.  This pushes support to a larger momentum domain in the pion, \emph{i.e}., a smaller distance domain.  One might ask whether there is a specific aspect of the data-driven pion GPD that is responsible.  The answer is ``yes'': EHM induced broadening of the pion wave function.

\section{Conclusion}
Amongst known fundamental theories of natural phenomena, QCD is unique.  Its scale-free Lagrangian is expressed in terms of degrees-of-freedom that are not directly observable.  That Lagrangian's massless gauge bosons become massive, owing solely to the self interactions between them.  That mass enables a stable, infrared completion of QCD through the emergence of a running coupling that saturates at infrared momenta, being everywhere finite.  These two effects ensure that massless quarks become massive and combine to form both massless Nambu Goldstone bosons and otherwise massive hadrons.

These emergent features of QCD are expressed in every strong interaction observable and they can also be revealed through EHM interference with Nature's other source of mass, \emph{i.e}., the Higgs boson.  With science's growing investment in high-energy, high-luminosity facilities, one can realistically expect to gather data in the foreseeable future that will enable validation of the EHM paradigm.  This could prove QCD to be the first well-defined four-dimensional quantum field theory that has ever been contemplated.  Such progress may open doors that provide insights into physics beyond the Standard Model.

\medskip

%
{\small \noindent{\bf Acknowledgments}. This contribution is based on ideas and insights developed through collaborations with many people, to all of whom I am greatly indebted.
Work supported by
the  National Natural Science Foundation of China (grant no.\,12135007).
}


%
%
%
%
%

\end{document}